\documentclass[conference]{IEEEtran}
\IEEEoverridecommandlockouts

\usepackage[T1]{fontenc}
\usepackage{xurl}
\usepackage{cite}
\usepackage{amsmath,amssymb,amsfonts}
\usepackage{algorithm}
\usepackage{algorithmic}
\usepackage{graphicx}
\usepackage{tikz}
\usepackage{textcomp}
\usepackage{xcolor}
\usepackage{pgfplots}
\usepackage{bm}
\usepackage{booktabs}

\usepackage{subcaption}
\usepackage{pgfplotstable}
\pgfplotsset{compat=1.18}
\usepackage{standalone}
\pgfplotsset{table/search path={figures}}
\usepackage{newtxmath}
\usepackage{balance}

\usetikzlibrary{arrows.meta,calc,decorations.markings,positioning,shapes.geometric}

\newcommand{\R}{\textcolor{brown}}

\begin{document}

\onecolumn
\thispagestyle{empty}
\vspace*{\fill}
\begin{center}
	{\large This work has been submitted to the IEEE for possible publication.
	Copyright may be transferred without notice, after which this version may no
	longer be accessible.\par}
	\vspace{1.5em}
	Submitted to the 2026 IEEE Future Networks World Forum (FNWF).
\end{center}
\vspace*{\fill}
\clearpage
\twocolumn
	
	\title{Opportunistic Positioning with LEO Satellites\\based on SSB from NR NTN}
	
	\author{\IEEEauthorblockN{Rainer Bachl\IEEEauthorrefmark{1}, Muhammad Nabeel\IEEEauthorrefmark{1}, Tingting Lei\IEEEauthorrefmark{1}\IEEEauthorrefmark{2}}
		
		\IEEEauthorblockA{\IEEEauthorrefmark{1}{Huawei Heisenberg Research Center, Munich, Germany}}
		\IEEEauthorblockA{\IEEEauthorrefmark{2}{Institute of Astronomical and Physical Geodesy, Technical University of Munich, Munich, Germany}}
		\IEEEauthorblockA{\texttt{Email: \{rainer.bachl,muhammad.nabeel8,tingting.lei1\}@huawei.com
		}}
	}
	
	\definecolor{matlab1}{RGB}{0, 114, 189}      
	\definecolor{matlab2}{RGB}{217, 83, 25}      
	\definecolor{matlab3}{RGB}{237, 177, 32}     
	\definecolor{matlab4}{RGB}{126, 47, 142}     
	\definecolor{matlab5}{RGB}{119, 172, 48}     
	
	\maketitle
	
	\begin{abstract}
		Forthcoming Low Earth Orbit (LEO) satellite networks such as Starlink's Mobile Satellite Service (MSS) will incorporate the New Radio (NR) Non-Terrestrial Network (NTN) standard.
		The Synchronization Signal Block (SSB) specified as part of NR is periodically broadcast for cell search and initial access. 
		We propose to exploit the SSB for opportunistic receiver positioning.
		Doppler shift measurements are modeled and pseudoranges are derived from SSB while also taking into account the receiver's clock bias and drift. The resulting per-satellite integer ambiguity in the pseudorange is resolved by geometry alone, without inter-satellite differencing or an a-priori position.
		Measurements are taken from SSBs of multiple satellites and at multiple occasions per satellite, whereby the SSBs are subject to different transmission timings and varying propagation delays.
		Finally, a simulation model is developed for positioning based on the actual Starlink constellation and the NR NTN standard to evaluate the positioning accuracy to be expected.
		The proposed approach achieves a mean positioning error of less than 10\,m without requiring any modification of the NR NTN standard.     
	\end{abstract}
	\begin{IEEEkeywords}
			LEO satellites, Starlink, Synchronization Signal Block, SSB, 3GPP NR NTN, opportunistic navigation, integer ambiguity resolution.
	\end{IEEEkeywords}
	
	\section{Introduction}
	\label{sec:intro}
	
	Several Low Earth Orbit (LEO) mega-constellations are currently planned or already deployed.
	Among them, Starlink has the densest constellation with more than 10\,000 satellites in orbit as of June 2026, and plans to extend this constellation further up to 15\,000 satellites as approved by the Federal Communications Commission (FCC)~\cite{fcc2026authorization}.
	With dense LEO constellations, a ground user has visibility of more satellites compared to the number of satellites visible from the Global Navigation Satellite System (GNSS)~\cite{morales2020gdop}.
	This makes LEO satellite based positioning attractive as it translates into a several-fold improvement over GNSS in geometric diversity for navigation~\cite{reid2018broadband}.
	Moreover, reduced propagation delays and higher received power from LEO satellites due to lower orbit heights highlight the potential of LEO systems in complementing existing GNSS, which is highly susceptible to jamming and spoofing~\cite{stock2024survey}.
	
	Currently operational LEO satellite systems focus exclusively on data communications and do not support receiver positioning.
	In particular, these LEO satellite systems employ proprietary standards which do not include recent satellite ephemeris signaling, and there is no transparent means for pseudorange measurements.
	Therefore, opportunistic LEO satellite based positioning currently suffers from low accuracy.
	The 3rd Generation Partnership Project (3GPP) has augmented the New Radio (NR) standards starting with Release~17 for Non-Terrestrial Networks (NTN)~\cite{tr38821}.
	This includes transparent signaling of recent and accurate satellite ephemeris.
	Starlink has committed in FCC submissions to setting up its Mobile Satellite Service (MSS) that will support NR NTN for Direct-To-Cell (DTC) communications starting in 2027~\cite{fcc2025mss}. 
	
	Within the scope of NR NTN, several works have recently focused on LEO satellite based positioning~\cite{gonzalezgarrido2024interference,dureppagari2024leo,zhu2024timing,edjekouane2025uere}.
	However, most of the literature relies on Positioning Reference Signals (PRS), which are not guaranteed to be available for positioning~\cite{gonzalezgarrido2024interference,dureppagari2024leo}.
	In~\cite{zhu2024timing}, a timing advance estimation method is proposed for receiver positioning using SSB signals from LEO satellites,
	and a comparison of PRS and SSB for Time of Arrival (ToA) based positioning in LEO networks is provided in~\cite{edjekouane2025uere} without considering Doppler measurements.
	Finally, authors of~\cite{neinavaie2021acquisition} focus on Starlink LEO satellites and achieve a horizontal positioning error of 10\,m by tracking Doppler only.
	However, since this approach must propagate satellite positions over long periods in the absence of recent ephemeris, the predicted positions may not be accurate.   
	
	In contrast to existing literature, we focus on receiver positioning based on SSB by making use of both Doppler shift and pseudorange in LEO satellite networks supporting NR NTN.
	To the best of the authors' knowledge, this is the first work which makes explicit use of NR NTN specifics to obtain recent ephemeris and derive pseudorange estimates for LEO satellite based positioning.
	
	Our main contributions can be summarized as follows:
	\begin{itemize}
		\item We demonstrate how to obtain the satellite ephemeris at the SSB transmit timing in NR NTN when the ground user is not synchronized to LEO satellites (Section~\ref{sec:ntn}).
		\item We derive pseudorange estimates, resolve the inherent per-satellite 10~ms integer ambiguity by geometry alone without inter-satellite differencing or an a-priori position, and use the pseudoranges jointly with Doppler measurements from SSB to obtain the receiver's position estimate (Section~\ref{sec:rx_processing}).
		\item Finally, we evaluate the accuracy of the proposed positioning approach by developing a realistic simulation model that exploits the Starlink constellation with NR~NTN (Section~\ref{sec:results}).
	\end{itemize}

	\section{Processing NR NTN Broadcast Signals}
	\label{sec:ntn}
	
	Throughout this work, we assume a LEO satellite constellation supporting NR NTN in which SSB is always present.
	Here, in Section~\ref{sec:ref_signals}, we first discuss the SSB and its benefits over other reference signals such as PRS.
	Next, for obtaining transmit timing information of the reference signals, a 10~ms radio-frame synchronized satellite network is assumed, which is discussed in Section~\ref{sec:sat-sync}.
	Finally, the actual computation of SSB transmit timing is outlined in Section~\ref{sec:tx_time}, and the computation of the ephemeris at the same timing instant is discussed in Section~\ref{sec:ephemeris}.
	
	\subsection{Reference Signals}
	\label{sec:ref_signals}
	
	In wireless communications, reference signals are used to obtain receive timing $T_{\text{Rx}}$ and frequency offset $\Delta f$ measurements.
	After downconversion to baseband, the process involves a two-dimensional correlation search over time and frequency with the expected reference signal waveforms, where the wide Doppler span of LEO satellites enlarges the frequency dimension of the search~\cite{yu2025ntnpss}. 
	Two families of downlink signals are candidates for these measurements on an NR NTN air interface: the dedicated PRS introduced for terrestrial NR positioning and extended towards NTN in 3GPP Release~18, and the always-on SSB. Several authors have proposed and analyzed PRS as the basis for NTN positioning,
	addressing PRS signal design, the accommodation of reference signals from multiple satellites within the communication resource grid, and interference for LEO deployments~\cite{gonzalezgarrido2024interference,dureppagari2024leo}, as well as carrier-phase enhancements~\cite{dureppagari2026carrierphase}.
	
	In this work, we focus on a positioning solution based on SSB, which is the more attractive choice for an opportunistic positioning receiver for four reasons. 
	First, the SSB is transmitted unconditionally for cell search and initial access, hence, it is present at all times and requires no positioning-specific configuration, dedicated assistance data or scheduling.
	In contrast, PRS must be explicitly configured and its parameters are not guaranteed to be available for a non-connected or opportunistic receiver. 
	Second, the SSB has a short repetition period chosen from $T_{\mathrm{SSB}}\in \{5, 10, 20, 40, 80, 160\}$~ms, with a default of $20$\,ms~\cite{ts38213}.
	Therefore, measurement opportunities recur on a fine, regular grid without waiting for a sparsely scheduled PRS occasion.
	Third, each SSB defines one of the satellite beams, and the ground user receives the selection of SSBs corresponding to its beam location. 
	Fourth, the SSB provides a large number of resource elements
	including the Primary Synchronization Signal (PSS), the Physical Broadcast Channel (PBCH) including the common demodulation reference signal, and the Secondary Synchronization Signal (SSS)~\cite{ts38211}. 
	After successful decoding of PBCH, all resource elements are known at the receiver and can be exploited coherently for obtaining refined timing and frequency measurements.
	Further details on SSB acquisition can be found in~\cite{wang2021pss,tuninato2023synchronization}.
	
	\subsection{Satellite Frame Alignment}
	\label{sec:sat-sync}
	
	Each satellite serves a large number of beams and therefore transmits a large number of SSBs to illuminate all beams.
	One SSB is broadcast in each beam and conveys the Physical Cell Identity (PCI), the System Frame Number (SFN), the half-frame indicator $n_{hf}$ and an SSB index~$\beta$.
	A satellite using a single PCI may transmit at most $L_{\max}$ SSBs, all of which are confined to the
	first 10\,ms radio frame of the SSB repetition period~\cite{ts38213}.
	The value of $L_{\max}$ is band- and numerology-dependent, and equals four for the L- and S-band NTN configuration assumed in this paper.
	
	Since a satellite using only one PCI can serve at most $L_{\max}$ beams, which is well below the per-satellite beam counts of typical NR NTN payloads~\cite{tr38821}, two mechanisms are used in NR.
	First, spatial reuse repeats the same SSB time-frequency resources in spatially separated beams with minimal mutual interference.
	Second, the satellite operates multiple PCIs, each maintaining its own independent SFN counter and
	each placing the first radio frame of its $T_{\mathrm{SSB}}$ window at a different starting offset.
	Combining multiple PCIs with the half-frame indicator $n_{hf}$, which distinguishes the two 5\,ms
	half-frames of every radio frame, allows the same satellite to transmit $L_{\max}$ SSBs in every 5\,ms half-frame.
	Consequently, the number of beams served per $T_{\mathrm{SSB}}$ scales as the product of the spatial reuse factor, the number of PCIs in use, and $L_{\max}$, whereby the number of PCIs is upper bounded by $T_{\mathrm{SSB}}/5~\text{ms}$.
	
	When multiple satellites illuminate the same beam area, these satellites need to avoid interference among SSB transmissions.
	Since each satellite carries an onboard GNSS receiver, it knows its own ephemeris and disciplines its clock to GNSS time.
	SSB interference is thus avoided by aligning the satellites' radio frames to a common 10 ms time base while sharing the SSB transmission opportunities created by multiple PCIs corresponding to specific SFN counters.
	Therefore, the resources for transmitting SSBs are also shared among the satellites serving the same beam area.
	It is important to note that these choices are compatible with, but not mandated by, the 3GPP NR NTN specifications; the common 10~ms time base is the key enabler for the pseudorange ambiguity resolution in Section~\ref{sec:ambiguity}. 
	
	\subsection{SSB Transmit Time}
	\label{sec:tx_time}
	
	The SSB transmit time $T_{\text{Tx}}$ is needed for computing the pseudorange estimate and as a parameter for the orbit propagation to obtain the satellite ephemeris at the time of SSB transmission.   
	In the satellite clock timeframe, the transmit time instant of the $k$-th SSB for satellite~$i$, denoted by $T_{\text{Tx},i,k}$, is obtained as
	\begin{equation}
		T_{\text{Tx},i,k} = N_{\text{SFN},i}(k)\cdot 10\,\text{ms}
		+ n_{hf}\cdot 5\,\text{ms}
		+ \sigma_\mu(\beta)\cdot T_{\text{sym}}^{(\mu)},
		\label{eq:t-tx-prop}
	\end{equation}
	where $N_{\text{SFN},i}(k)$ is the SFN counter of satellite~$i$ for the $k$-th SSB and $n_{hf}$ is the half-frame indicator, both decoded from PBCH of SSB.
	Moreover, $\sigma_\mu(\beta)$ denotes the starting OFDM-symbol index of SSB candidate $\beta$ within the half-frame for sub-carrier-spacing numerology $\mu$, given by the SSB candidate-position table in~\cite{ts38213}, with $T_{\text{sym}}^{(\mu)}$ as the corresponding OFDM symbol duration.
	
	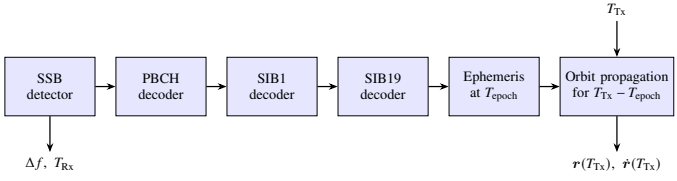
\begin{figure}[t]
		\centering
		\resizebox{\columnwidth}{!}{%
			\begin{tikzpicture}[
				block/.style={rectangle, draw, fill=blue!10,
					minimum width=2.0cm, minimum height=1.3cm,
					align=center, font=\small},
				io/.style={font=\small},
				arrow/.style={-{Stealth[length=2mm]}, thick},
				node distance=0.45cm
				]
				\node[block]                          (b1) {SSB\\detector};
				\node[block, right=of b1]             (b2) {PBCH\\decoder};
				\node[block, right=of b2]             (b3) {SIB1\\decoder};
				\node[block, right=of b3]             (b4) {SIB19\\decoder};
				\node[block, right=of b4]             (b6)
				{Ephemeris\\at $T_{\text{epoch}}$};
				\node[block, right=of b6]             (b7)
				{Orbit propagation\\for $T_{\text{Tx}} - T_{\text{epoch}}$};
				
				\draw[arrow] (b1) -- (b2);
				\draw[arrow] (b2) -- (b3);
				\draw[arrow] (b3) -- (b4);
				\draw[arrow] (b4) -- (b6);
				\draw[arrow] (b6) -- (b7);
				
				\node[io, below=0.8cm of b1]       (out1) {$\Delta f,\ T_{\text{Rx}}$};
				\draw[arrow] (b1) -- (out1);
				
				\node[io, above=0.8cm of b7]       (in7) {$T_{\text{Tx}}$};
				\draw[arrow] (in7) -- (b7);
				
				\node[io, below=0.8cm of b7]       (out7)
				{$\bm{r}(T_{\text{Tx}}),\ \dot{\bm{r}}(T_{\text{Tx}})$};
				\draw[arrow] (b7) -- (out7);
			\end{tikzpicture}%
		}
		\caption{Serving cell SSB processing for positioning.
			The orbit propagator uses a computed $T_{\text{Tx}}$ and yields the satellite state
			$(\bm{r}, \dot{\bm{r}})$ valid at the SSB transmission instant.}
		\label{fig:ssb-serving-cell}
	\end{figure}
	
	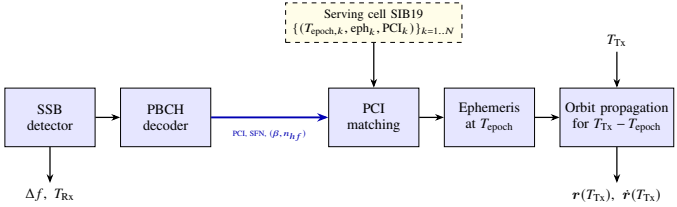
\begin{figure}[t]
		\centering
		\resizebox{\columnwidth}{!}{%
			\begin{tikzpicture}[
				block/.style={rectangle, draw, fill=blue!10,
					minimum width=2.0cm, minimum height=1.3cm,
					align=center, font=\small},
				io/.style={font=\small},
				ext/.style={rectangle, draw, dashed, fill=yellow!10,
					align=center, font=\footnotesize, inner sep=4pt},
				arrow/.style={-{Stealth[length=2mm]}, thick},
				node distance=0.55cm
				]
				\node[block]                          (b1) {SSB\\detector};
				\node[block, right=of b1]             (b2) {PBCH\\decoder};
				\node[block, right=2.6cm of b2]       (b3) {PCI\\matching};
				\node[block, right=of b3]             (b4)
				{Ephemeris\\at $T_{\text{epoch}}$};
				\node[block, right=of b4]             (b5)
				{Orbit propagation\\for $T_{\text{Tx}} - T_{\text{epoch}}$};
				
				\draw[arrow] (b1) -- (b2);
				\draw[-{Stealth[length=2.5mm]}, very thick, blue!70!black]
				(b2) -- (b3)
				node[midway, below=4pt, font=\tiny, blue!70!black]
				{PCI, SFN, $(\beta,n_{hf})$};
				\draw[arrow] (b3) -- (b4);
				\draw[arrow] (b4) -- (b5);
				
				\node[io, below=0.8cm of b1]        (out1) {$\Delta f,\ T_{\text{Rx}}$};
				\draw[arrow] (b1) -- (out1);
				
				\node[ext, above=1.0cm of b3]       (sib)
				{Serving cell SIB19\\
					$\{(T_{\text{epoch},k}, \text{eph}_k, \text{PCI}_k)\}_{k=1..N}$};
				\draw[arrow] (sib) -- (b3);
				
				\node[io, above=0.8cm of b5]        (in5) {$T_{\text{Tx}}$};
				\draw[arrow] (in5) -- (b5);
				
				\node[io, below=0.8cm of b5]        (out5)
				{$\bm{r}(T_{\text{Tx}}),\ \dot{\bm{r}}(T_{\text{Tx}})$};
				\draw[arrow] (b5) -- (out5);
			\end{tikzpicture}%
		}
		\caption{SSB processing when the ephemeris is known from neighbor cell information with prior serving cell SIB19 decoding. The neighbor cell's  $(\text{ephemeris}, T_{\text{epoch}})$ pair is supplied by the previously-decoded serving cell's
			SIB19 (dashed box).
			PCI-matching identifies the ephemeris of the satellite that has transmitted the SSB.}
		\label{fig:ssb-neighbor-cell}
	\end{figure}
	
	\subsection{Ephemeris and Orbit Propagation}
	\label{sec:ephemeris}
	
	The ephemeris of the serving satellite is broadcast through System Information Block 19 (SIB19)~\cite{ts38331}.
	The scheduling information for SIB19 is provided in SIB1, which is acquired after decoding the PBCH available in the SSB.
	The complete procedure is depicted in Fig.~\ref{fig:ssb-serving-cell}.
	The same SIB19 also includes the ephemerides of neighboring satellites, along with their PCIs and the corresponding reference epoch $T_{\text{epoch}}$ expressed in terms of SFN and subframe number.
	SIB19 is broadcast periodically, with intervals as short as a few seconds, and the reference epoch $T_{\text{epoch}}$ is based on the satellite's onboard GNSS solution.
	The SSB detection of neighboring satellites works in the same way as for the serving satellite.
	However, there is no need to decode the ephemeris of neighboring satellites as the information is already derived from the serving satellite as depicted in Fig.~\ref{fig:ssb-neighbor-cell}.
	The orbit propagator shown as a last step in Figs.~\ref{fig:ssb-serving-cell} and~\ref{fig:ssb-neighbor-cell} requires two inputs, i.e., ephemeris information together with its $T_{\text{epoch}}$ and the SSB transmit instant $T_{\text{Tx}}$.
	Both time instants are expressed in terms of the same SFN with respect to the same satellite clock.
	
	Orbit propagation can then be used to advance the received ephemeris from $T_{\text{epoch}}$ to $T_{\text{Tx},i,k}$ and to obtain the satellite state in ECEF coordinates for position and velocity, $(\bm{r}_i(T_{\text{Tx},i,k}),\dot{\bm{r}}_i(T_{\text{Tx},i,k}))$, at the SSB transmit time instant.
	Commonly used orbit propagators include SGP4 when the ephemeris is given in Two-Line-Element (TLE) form~\cite{vallado2006sgp4}, or a plain two-body Keplerian propagator when it is given as a set of orbital elements~\cite{vallado2013astrodynamics,nabeel2025low}.
	Since $T_{\text{epoch}}$ is only seconds ahead of its use, position and velocity errors produced by these orbit propagators are negligible in comparison to SSB measurement noise that dominates the positioning budget.
	Therefore, for the remainder of this paper, it is assumed that the orbit propagation output $(\bm{r}_i(T_{\text{Tx},i,k}),\dot{\bm{r}}_i(T_{\text{Tx},i,k}))$ corresponds to the correct satellite ephemeris at the SSB transmit instant.

\section{Measurements and Positioning}
\label{sec:rx_processing}

\subsection{Doppler Measurement}
\label{sec:doppler-meas}

The Doppler measurement is the deviation of the measured carrier frequency from its nominal value $f_c$. 
For the $k$-th SSB of satellite~$i$ it takes the form
\begin{equation}
	\Delta f_{i,k} = f^{d}_{i,k} + d\,f_c + \nu^{f}_{i,k},
	\label{eq:doppler-meas}
\end{equation}
where $f^{d}_{i,k}$ is the geometric Doppler shift caused by the relative motion between satellite $i$ and the ground user, 
$d$ is the receiver clock drift (in seconds per second, shared across all satellites and occasions), 
$d\,f_c$ is the resulting frequency offset, 
and $\nu^{f}_{i,k}$ is zero-mean Gaussian measurement noise.
The geometric Doppler $f^{d}_{i,k}$ depends on the satellite-$i$ ephemeris delivered by the orbit propagator described in Section~\ref{sec:ephemeris} and on the unknown ground user position through the standard line-of-sight relation~\cite{psiaki2021transit}.
Over the short batch windows considered in this work, the receiver clock drift $d$ is assumed constant and is estimated jointly with the ground user position. 
3GPP TS 38.101-5 additionally bounds the magnitude of $d$: NTN receivers should keep the modulated carrier frequency within $\pm 0.1$~ppm of the ideally Doppler-pre-compensated reference~\cite{ts38101_5}, giving $|d|\le 10^{-7}~\text{s/s}$. While this requirement governs the UE transmitter, we adopt it as representative of the oscillator quality of an NTN-capable device.

\subsection{Pseudorange Measurement}
\label{sec:pseudorange-meas}

The pseudorange measurement for the $k$-th SSB of satellite~$i$ is constructed as $\mathrm{PR}_{i,k} = c\,(T_{\text{Rx},i,k} - T_{\text{Tx},i,k})$, 
where $T_{\text{Rx},i,k}$ is given by the SSB detector and $T_{\text{Tx},i,k}$ is obtained from \eqref{eq:t-tx-prop} using the SIB19 ephemeris. 
The per-satellite time reference $T_{0,i}$, defined as the network time
at which the current SFN cycle of satellite~$i$ begins, places
$T_{\text{Tx},i,k}$ in a common network time frame; it is not signaled
in any field of SIB19. The receiver timestamps the SSB arrival with its
local clock, which deviates from network time by the bias $b$ and
drift $d$, so that
\begin{equation}
	T_{\text{Rx},i,k}
	= T_{0,i} + T_{\text{Tx},i,k} + \frac{\rho_{i,k}}{c}
	+ b + d\,t_{i,k} + \frac{\nu^{\mathrm{pr}}_{i,k}}{c},
	\label{eq:rx-time-model}
\end{equation}
where $t_{i,k}$ is the measurement time relative to the batch origin,
$\rho_{i,k}$ is the geometric range between the ground user and the satellite, 
$c$ is the speed of light, 
and $\nu^{\mathrm{pr}}_{i,k}$ is zero-mean Gaussian measurement noise.
Under the frame-aligned, GNSS-synchronized deployment of
Section~\ref{sec:sat-sync}, every $T_{0,i}$ lies on a common 10~ms
lattice, i.e., $T_{0,i} = K_i\cdot 10~\text{ms}$ with
$K_i\in\mathbb{Z}$ a per-satellite integer offset that also absorbs
the SFN wrap-around. Substituting~\eqref{eq:rx-time-model} into the pseudorange definition yields
\begin{equation}
    \mathrm{PR}_{i,k} 
    = \rho_{i,k} + c\,b + c\,d\,t_{i,k} + c\,K_i\cdot 10~\text{ms} + \nu^{\mathrm{pr}}_{i,k}.
    \label{eq:pseudorange-meas}
\end{equation}

\subsection{Integer Ambiguity Resolution}
\label{sec:ambiguity}

The integer $K_i$ remains as a per-satellite ambiguity of magnitude $c\,K_i\cdot 10~\text{ms}\approx K_i\cdot 2998$~km in $\mathrm{PR}_{i,k}$, and for a LEO constellation it is resolved by geometry alone, without any extra signaling. 
Since $K_i$ is per-satellite, the occasion index $k$ is omitted in $\rho_i$, $\mathrm{PR}_i$ and $t_i$ in the geometry-only resolution described in the remainder of this subsection. 
With the ground user on the Earth's surface, satellite altitude $h\in[300,1500]$~km and a worst-case elevation limit $\varepsilon_{\min}=10^{\circ}$ (the simulations in Section~\ref{sec:results} use $\varepsilon_{\min}=30^{\circ}$), the geometric range is physically bounded by $\rho_i \in [h,\,\rho_{\max}]$, where the maximum slant range is
\begin{equation}
    \rho_{\max} = \sqrt{(R_E+h)^2 - R_E^2\cos^2\varepsilon_{\min}}
        - R_E\sin\varepsilon_{\min},
    \label{eq:leo-range-bracket}
\end{equation}
with $R_E$ the Earth radius. 
At the top of the LEO altitude band $\rho_{\max}$ is about $3650$~km, so the propagation delay $\tau_i = \rho_i/c$ lies in roughly $[1,13]$~ms -- a span comparable to one ambiguity step.


The known orbital altitude resolves the ambiguity without an a-priori ground user position. 
The ambiguity is resolved from one occasion per satellite within the
first SSB period, so that $|t_i| \le T_{\mathrm{SSB}}$ holds and the
drift term contributes at most
$c\,|d|\,|t_i| \le c\cdot 10^{-7}\cdot T_{\mathrm{SSB}} < 5$~m, which
is negligible against the ambiguity step and therefore neglected in
the following. Defining
$\varphi \equiv b \bmod 10~\text{ms}\in[0,10~\text{ms})$, the implied
geometric range for a trial $(\varphi,K_i)$ is
\begin{equation}
    \rho_i(\varphi,K_i) = \mathrm{PR}_i - c\,\varphi
        - c\,K_i\cdot 10~\text{ms}.
    \label{eq:rho-implied}
\end{equation}
Only $\varphi$ and the integers $K_i$ are separable here: the integer
part of $b/10~\text{ms}$ shifts all $K_i$ by a common amount and is
absorbed into them, so the trial integers represent the per-satellite
offsets up to this common shift.
The true pair must respect the physical altitude bracket
\begin{equation}
    h_i \le \rho_i(\varphi,K_i) \le \rho_{\max}(h_i).
    \label{eq:rho-bracket-constraint}
\end{equation}
Since $\rho_{\max}(h_i) - h_i < c\cdot 10~\text{ms}$, the bracket is
only $3$--$7$~ms wide across the LEO band and thus narrower than one
ambiguity step (this inequality first fails at $h_i \approx 3400$~km, far above the shells considered here). Consequently, at most one integer $K_i$
satisfies~\eqref{eq:rho-bracket-constraint} for a given $\varphi$.
Each satellite therefore constrains $\varphi$ to a single interval $\Phi_i\subset[0,10~\text{ms})$ of width $|\Phi_i| = \left(\rho_{\max}(h_i) - h_i\right)/c$, on which the integer $K_i(\varphi)$ is constant. 
The true phase lies in the intersection $\varphi \in \bigcap_i \Phi_i$, which collapses to a narrow sub-interval of $[0,10~\text{ms})$ for a handful of satellites with diverse geometry. 
Since each $K_i(\varphi)$ is constant on its own $\Phi_i$, any $\varphi$ in the non-empty intersection fixes the whole integer set $\{K_i\}$. 
The residual width of $\bigcap_i\Phi_i$ is merely a coarse estimate of $b \bmod 10~\text{ms}$, refined together with the position by the solver of Section~\ref{sec:positioning}. In the simulations of Section~\ref{sec:results}, the integer ambiguities were resolved correctly in all trials using the first occasion of the $N\le 8$ selected satellites.

\subsection{Positioning}
\label{sec:positioning}
A GNSS-less receiver obtains pseudorange and Doppler measurements from SSB signals of multiple selected LEO satellites that are synchronized to a common network time reference. 
Assuming a static ground user and neglecting channel distortions, the vector of unknowns is $\mathbf{x} = [\mathbf{r}^{T}\;\, b\;\, d]^{T}$, where $\mathbf{r}$ is the ground user position, $b$ the receiver clock bias (modulo 10 ms, see Section~\ref{sec:ambiguity}), and $d$ the receiver clock drift.

With satellites $i=1,\ldots,N$ selected for positioning and the SSB occasions of satellite $i$ given by $k(i)=1,2,\ldots,M_i$, each pair $(i,k)$ denotes one SSB measurement. 
The pseudorange and Doppler measurement equations~\eqref{eq:pseudorange-meas} and~\eqref{eq:doppler-meas} relate each measurement to the unknowns $\mathbf{x}$ up to measurement noise. 
After resolving the per-satellite integer ambiguity $K_i$ by geometry (Section~\ref{sec:ambiguity}), positioning proceeds with the ambiguity-corrected pseudorange
\begin{equation}
	\tilde{\mathrm{PR}}_{i,k} = \mathrm{PR}_{i,k} - c\,K_i\cdot 10~\text{ms}.
	\label{eq:pr-corrected}
\end{equation}
The clock-bias unknown estimated in the sequel is thus
$b \bmod 10~\text{ms}$, the absolute integer multiple having been
absorbed into the resolved $K_i$.
The deterministic measurement models are
\begin{equation}
	\begin{aligned}
		h^{\rho}_{i,k}(\mathbf{x}) &= \rho_{i,k}(\mathbf{r}) + c\,b + c\,d\,t_{i,k},\\
		h^{f}_{i,k}(\mathbf{x}) &= f^{d}_{i,k}(\mathbf{r}) + d\,f_c,
	\end{aligned}
	\label{eq:pos-models}
\end{equation}
where the geometric range $\rho_{i,k}(\mathbf{r})$ and Doppler shift $f^{d}_{i,k}(\mathbf{r})$ depend on the ground user position $\mathbf{r}$ only and are evaluated with the ephemeris of satellite $i$ at the transmit instant $T_{\mathrm{Tx},i,k}$ of its $k$-th SSB. 
Because SSB schedules and propagation delays differ across satellites and occasions, the geometry is specific to each $(i,k)$ rather than referred to a single receiver epoch.

The ground user state is estimated by minimizing the weighted least-squares cost function
\begin{equation}
	\begin{aligned}
		\hat{\mathbf{x}} &= \arg\min_{\mathbf{x}}\, J(\mathbf{x}),\\
		J(\mathbf{x}) &= \sum_{i}\sum_{k(i)}\frac{\bigl(\tilde{\mathrm{PR}}_{i,k}-h^{\rho}_{i,k}(\mathbf{x})\bigr)^2}{\sigma_{\rho}^2}\\
		&\quad+ \sum_{i}\sum_{k(i)}\frac{\bigl(\Delta f_{i,k}-h^{f}_{i,k}(\mathbf{x})\bigr)^2}{\sigma_{f}^2},
	\end{aligned}
	\label{eq:hybrid_wls}
\end{equation}
where $\sigma_{\rho}$ and $\sigma_{f}$ are the pseudorange and Doppler standard deviations, the outer sum runs over the $N$ selected satellites, and the inner sum over the $M_i$ SSB occasions of satellite $i$. Each squared residual is normalized by the variance of its own measurement type, so $\sigma_{\rho}$ and $\sigma_{f}$ must be expressed in the units of the corresponding residual (meters and Hertz, respectively); this normalization renders the two heterogeneous terms dimensionless and commensurable.


The cost function~\eqref{eq:hybrid_wls} is nonlinear in $\mathbf{x}$ and is solved iteratively. 
We use the adaptive-weights damped Gauss--Newton method with backtracking (AW-DGN-BT) of~\cite{morichi2026convergence}, which is robust for the ill-conditioned geometries typical of LEO satellite constellations. The solver is initialized from a coarse position drawn within the initialization error of Table~\ref{tab:simulation_param}, and no divergence was observed across the simulated trials.

\section{Results and Discussion}
\label{sec:results}

We focus on positioning with a realistic DTC satellite constellation for ground users using future NR NTN cellphones.
Currently, there is no such satellite constellation available and the Starlink DTC constellation is still incomplete. Therefore, we consider two constellation setups with (i) all current 9\,903 non-DTC Starlink satellites as of June 2026, and (ii) a constellation of 12\,000 satellites with orbital shells selected from SpaceX's pending Starlink MSS application~\cite{fcc2025space}, but otherwise similar to the non-DTC Starlink constellation.  

\subsection{Simulation Setup}
\label{sec:sim_setup}

\begin{figure}[t]
	\centering
	\includegraphics[width=\columnwidth]{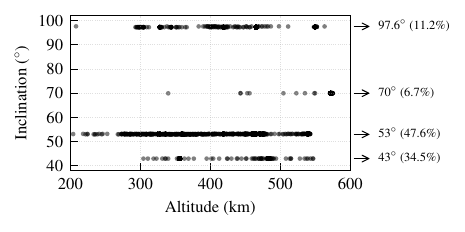}
	\caption{Inclination--altitude distribution of the current non-DTC Starlink constellation.}
	\label{fig:starlink_distribution}
\end{figure}

The TLEs for non-DTC Starlink satellites are extracted from CelesTrak\footnote{\url{https://celestrak.org/}} and used as a first constellation setup for simulation.
The distribution of orbital inclinations and altitudes of current non-DTC Starlink constellation is shown in Fig.~\ref{fig:starlink_distribution}.
Using a similar setup, we model a second constellation based on Starlink MSS with 4 orbital shells, each in a Walker delta configuration. 
The inclinations, the percentage of overall satellites and the orbital planes per orbital shell are chosen to be similar to the current Starlink non-DTC constellation but consistent with SpaceX's pending Starlink MSS application to the FCC~\cite{fcc2025space}.
SpaceX's pending MSS application requests up to 15\,000 satellites~\cite{fcc2025space}.
We nevertheless model only 12\,000 satellites (still more than the current Starlink non-DTC constellation), since their lower orbit altitudes below 335\,km imply smaller per-satellite coverage areas.
Detailed configuration of the created constellation is provided in Table~\ref{tab:synthetic_constellation_configuration}. 

The satellite ephemerides are propagated with SGP4 only to the times of SSB transmissions.
Ephemeris errors are neglected, as justified in Section~\ref{sec:ephemeris}. 
The SSB transmit timings are chosen compliant with NR, but fixed for a particular beam representing an Earth-fixed beam configuration.
The SSB transmit timings are distinct per satellite, and satellites selected for illuminating a beam occupy one of the SSB transmission opportunities; when a satellite vanishes from visibility for a beam, its opportunities are released and taken over by the next satellite chosen to illuminate that beam. 

An elevation angle above $30^\circ$ is used as the satellite visibility threshold.
Over the simulation period, an average of 53 and 27 satellites are visible at the considered receiver location for the non-DTC Starlink and Starlink MSS constellations, respectively.
The lower number of visible satellites in the Starlink MSS constellation is due to its lower orbital altitude.
Nevertheless, the minimum number of visible satellites at any time is 40 and 22 for the non-DTC Starlink and Starlink MSS constellations, respectively.
Among visible satellites, we choose up to 8 satellites for providing coverage in that beam based on the largest positive Doppler shift, i.e., the rising satellites among the visible ones. 
Selecting rising satellites also guarantees the longest remaining visibility period for the chosen satellites.

Each reported mean positioning error is obtained from 400 estimation trials, where the trials differ only in their realizations of measurement noise, clock error, and initialization. 
Using a fixed random seed, the same 400 realizations of noise, clock error, and initialization are used for every configuration, so that any two configurations are evaluated on identical conditions.
The start times along the orbital pass are also fixed across experiments, ensuring all configurations share a common starting point.
Other simulation parameters are summarized in Table~\ref{tab:simulation_param}.
The standard deviations of normally distributed noise in frequency offset and pseudorange measurements are assumed identical across satellites and occasions.
We have ignored tropospheric and ionospheric channel distortions. At 2\,GHz and $30^\circ$ elevation these amount to roughly 5~m of tropospheric and 1--10\,m of ionospheric delay, comparable to $\sigma_\rho$, and could be compensated upfront based on coarse knowledge of the ground user location, leaving a sub-noise residual. The simulated clock bias is drawn up to $10^{-6}$\,s (see Table~\ref{tab:simulation_param}); this is without loss of generality, since the integer part of $b/10\,\text{ms}$ is absorbed into the $K_i$ (described in Section~\ref{sec:ambiguity}) and only the fractional part affects positioning.

\begin{table}[t]
	\centering
	\caption{Constellation based on Starlink MSS}
	\setlength{\tabcolsep}{2pt}
	\begin{tabular*}{\columnwidth}{@{\hspace{2pt}\extracolsep{\fill}}ccccccc@{\hspace{2pt}}}
		\toprule
		\textbf{Orbital} & \textbf{Inclination} & \textbf{Altitude} & \textbf{Planes} & \textbf{Sats per} & \textbf{Total} & \textbf{Percentage} \\
		\textbf{Shell}   & \textbf{($^\circ$)}  & \textbf{(km)}     &                  & \textbf{Plane}     & \textbf{Sats} &        \\
		\midrule
		1 & $43$    & 332 & 68 & 60 & 4080 & 34\% \\
		2 & $53$    & 330 & 96 & 60 & 5760 & 48\% \\
		3 & $69$    & 328 & 28 & 30 &  840 &  7\% \\
		4 & $96.87$ & 326 & 22 & 60 & 1320 & 11\% \\
		\midrule
		\multicolumn{5}{r}{\textbf{Total}} & \textbf{12000} & \textbf{100\%} \\
		\bottomrule
	\end{tabular*}
	\label{tab:synthetic_constellation_configuration}
\end{table} 

\begin{table*}[t]
	\centering
	\caption{Simulation Parameters}
	\begin{tabular}{ll@{\hspace{20pt}}ll}
		\toprule
		\textbf{Parameter} & \textbf{Value} & \textbf{Parameter} & \textbf{Value} \\
		\midrule
		Carrier frequency & $f_c = 2$ GHz &
		Max. sats providing beam coverage & 8 \\
		SSB period & 160 ms &
		Min. elevation for visibility & 30\textdegree \\
		Beams per satellite & 128 &
		Max. clock bias (uniform dist.) & $10^{-6}$ s \\
		Subcarrier spacing & 30 kHz &
		Max. clock drift (uniform dist.)& $10^{-7}$ s/s \\
		Ground user location & Munich (48.14\textdegree N, 11.58\textdegree E) &
		Std. dev. init. pos. error (normal dist.) & 100~km \\
		Sat. selection criterion & Max. positive Doppler &
		Number of trials per pos. error result &  400 \\ 		
		\bottomrule
	\end{tabular}
	\label{tab:simulation_param}
\end{table*}

\subsection{Simulation Results}

\begin{figure}[t]
	\centering
	\begin{subfigure}[b]{\columnwidth}
		\centering
		\includegraphics[width=0.8\columnwidth]{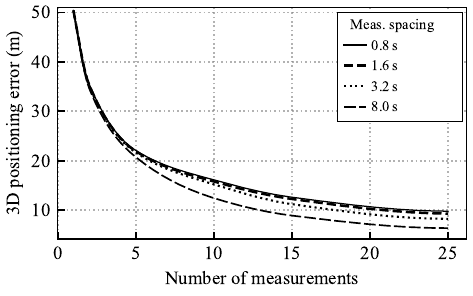}
		\caption{Non-DTC Starlink constellation.}
		\label{fig:starlink_measurement_distance}
	\end{subfigure}
	\hfill
	\begin{subfigure}[b]{\columnwidth}
		\centering
		\includegraphics[width=0.8\columnwidth]{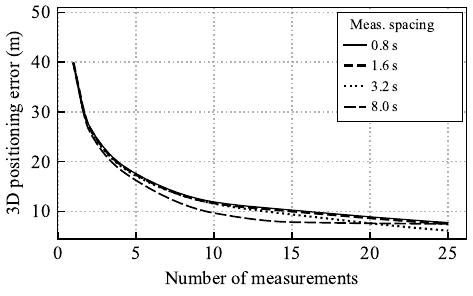}
		\caption{Constellation based on Starlink MSS.}
		\label{fig:synthetic_measurement_distance}
	\end{subfigure}
	\caption{Mean positioning error, with standard deviations of the measurement error of 100\,Hz for Doppler and 10\,m for pseudorange.}
	\label{fig:measurement_distance_combined}
\end{figure}

For both satellite constellations considered, we investigate the positioning accuracy by varying the measurement set used to obtain one positioning result. 
In particular, we vary the number of measurements and the spacing between consecutive measurements in the measurement set. 
Then up to 8 satellites providing beam coverage at the start of each set as described previously are used for the measurements. 
The satellites are included in the measurement set as long as the number of measurements considered is not exceeded, and as long as they are visible to the ground user based on the minimum elevation criterion. 
In the simulation, the standard deviation of the measurement errors is chosen as 100\,Hz for Doppler and 10\,m for pseudorange~\cite{edjekouane2025uere}.
The mean 3D positioning errors for non-DTC Starlink and the constellation based on Starlink MSS are plotted in Fig.~\ref{fig:measurement_distance_combined} for measurement spacings of 0.8\,s, 1.6\,s, 3.2\,s and 8\,s, which are multiples of the SSB periodicity.

As seen in Fig.~\ref{fig:measurement_distance_combined}, more measurements and larger measurement spacings yield better positioning accuracy in most cases. 
A larger number of measurements helps to average out the measurement errors. 
A larger measurement spacing yields a better geometric diversity within the measurement set.
However, a larger measurement spacing also increases the overall time needed for positioning.
For the longest measurement spacings, some satellites even vanish from visibility during the measurement period (which can be seen by the flattening out of the 3D positioning error curve vs. number of measurements for measurement spacing of 8\,s in Fig.~\ref{fig:measurement_distance_combined}b).

\begin{figure}[t]
	\centering
	\begin{subfigure}[b]{\columnwidth}
		\centering
		\includegraphics[width=0.8\columnwidth]{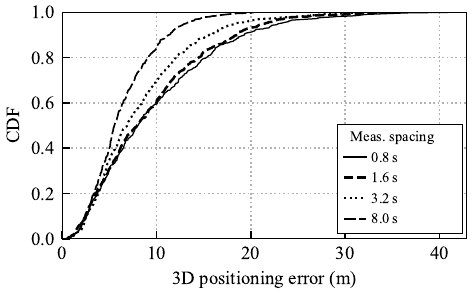}
		\caption{Non-DTC Starlink constellation.}
		\label{fig:starlink_cdf}
	\end{subfigure}
	\hfill
	\begin{subfigure}[b]{\columnwidth}
		\centering
		\includegraphics[width=0.8\columnwidth]{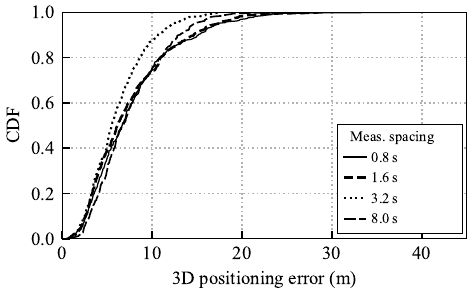}
		\caption{Constellation based on Starlink MSS.}
		\label{fig:synthetic_cdf}
	\end{subfigure}
	\caption{Positioning error CDF after 25 measurements; measurement noise as in Fig.~\ref{fig:measurement_distance_combined}.}
	\label{fig:cdf_combined}
\end{figure}

The distribution of the positioning error is illustrated with the CDF after 25 measurements in Fig.~\ref{fig:cdf_combined}.
The resulting error is less than 20\,m in 90\,\% of the positioning trials. For 25 measurements at a spacing of 3.2\,s, the mean 3D positioning error is 8.2\,m for the non-DTC Starlink constellation and 6.1\,m for the constellation based on Starlink MSS.
Similar to the previous observation, the performance statistics do not improve when choosing measurement spacings longer than 3.2\,s for the constellation based on Starlink MSS.

\begin{figure}[t]
	\centering
	\includegraphics[width=\columnwidth]{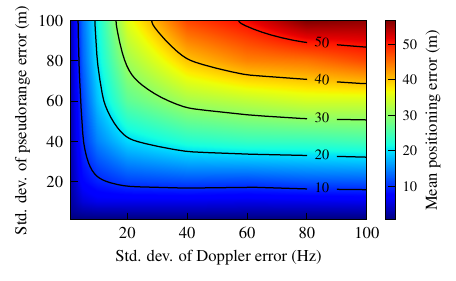}
	\caption{Mean positioning error as a function of Doppler and pseudorange measurement errors, with measurement spacing of 3.2\,s after 25 measurements.}
	\label{fig:synthetic_heatmap_zoom}
\end{figure}

Finally, the impact of Doppler and pseudorange measurement error on the positioning accuracy is shown in Fig.~\ref{fig:synthetic_heatmap_zoom} for the constellation based on Starlink MSS using 25 measurements with a measurement spacing of 3.2\,s. 
It highlights that the positioning accuracy is more sensitive to pseudorange measurement accuracy than to Doppler measurement accuracy, and accuracy is substantially improved through Doppler measurements only when the corresponding measurement errors are smaller than the practically achievable Doppler accuracy~\cite{duval2026towards}.
Nevertheless, in case the Doppler estimation is accurate, then regardless of pseudorange measurements, a good positioning accuracy can be achieved similar to as also demonstrated in~\cite{neinavaie2021acquisition}.
In summary, even with high errors in pseudorange and Doppler measurements, it is possible to achieve an accuracy of a few tens of meters by using only SSB, and without any modification in the existing NR NTN standard.

\section{Conclusion}
\label{sec:conclusions}
We investigated opportunistic positioning using the always-available SSB transmitted by LEO satellites with NR NTN. 
SSBs from different satellites, transmitted at different time instants, are jointly processed, from which pseudorange and Doppler measurements are extracted. 
Simulation results based on Starlink constellations show that positioning based on SSB achieves an accuracy of less than 10\,m under the considered assumptions. 
These results indicate that SSB signals can provide a practical and resource-efficient positioning opportunity in future dense LEO NR NTN deployments. 

\balance

\bibliographystyle{IEEEtran}
\bibliography{references}

\end{document}